 \documentclass[final,5p,times,twocolumn,numbers]{elsarticle}

\usepackage{amssymb}
\usepackage{lipsum}
\usepackage{amsmath}
\usepackage{physics}
\usepackage{tikz}
\usepackage{tikz-3dplot}
\usepackage{amsmath}
\usepackage{natbib}
\usetikzlibrary{angles,quotes}
\usetikzlibrary{arrows.meta}

\usepackage{textpos} % プレプリ番号表示

\usepackage[colorlinks=true, linkcolor=blue, citecolor=blue, urlcolor=blue]{hyperref}

\journal{Physics Letters B}

\begin{document}

\begin{frontmatter}



\title{Estimate of virtual photon polarization due to the intense magnetic field in Pb-Pb collisions at the LHC energies}


\author[hiroshima]{Kento Kimura}
\author[hiroshima]{Nicholas J. Benoit}
\author[hiroshima]{Ken-Ichi Ishikawa}
\author[hiroshima,skcm2,nagoya,kmi]{Chiho Nonaka}
\author[hiroshima,skcm2]{Kenta Shigaki}
\affiliation[hiroshima]{organization={Physics Program, Hiroshima University},
            addressline={}, 
            city={Higashi-Hiroshima},
            postcode={739-8526}, 
            state={Hiroshima},
            country={Japan}}
\affiliation[skcm2]{organization={International Institute for Sustainability with Knotted Chiral Meta Matter (WPI-SKCM$^2$), Hiroshima University},
            addressline={}, 
            city={Higashi-Hiroshima},
            postcode={739-8526}, 
            state={Hiroshima},
            country={Japan}}
\affiliation[nagoya]{organization={Department of Physics, Nagoya University},
            addressline={}, 
            city={Nagoya},
            postcode={464-8602}, 
            state={Aichi},
            country={Japan}}
\affiliation[kmi]{organization={Kobayashi-Maskawa Institute for the Origin of Particles and the Universe, Nagoya University},
            addressline={}, 
            city={Nagoya},
            postcode={464-8602}, 
            state={Aichi},
            country={Japan}}
            
\begin{abstract}
    We present the first numerical calculation of the virtual photon polarization and assess the feasibility of measuring the polarization via the anisotropic decay $\gamma^{*} \rightarrow \mu\mu$ using the LHC-ALICE detector.
    In presence of intense magnetic fields generated in high-energy non-central heavy-ion collisions that exceed the critical magnetic field intensity of quantum electrodynamics (QED), prompt virtual photons are predicted to decay anisotropically into lepton pairs, which we call virtual photon polarization.
    Using a relativistic resistive magnetohydrodynamics model, we computed the time evolution of the magnetic field and used these results to estimate the averaged polarization by calculating the vacuum polarization under the influence of the magnetic field at specific times.
    The estimated polarization deviates from zero with a statistical significance of $0.07\sigma$ with the data statistics collected from 2010 to 2011 and $0.15\sigma$ with the one from 2015 to 2018. It is understandable that the magnetic field could not be detected through polarization due to low statistical significance. With the data collecting the ongoing ALICE run from 2023 to 2026, the statistics dramatically increase by the upgraded LHC and the new data processing system. Thereby we expect that the statistical significance could reach $\sim  1.7\sigma$, resulting in a promising probe for detecting the intense magnetic fields.
\end{abstract}



\begin{keyword}
%% keywords here, in the form: keyword \sep keyword, up to a maximum of 6 keywords
Relativistic heavy-ion collisions \sep intense magnetic field \sep non-linear QED



\end{keyword}

\end{frontmatter}

%\tableofcontents


%% プレプリ番号表示
\setlength{\TPHorizModule}{1em}
\setlength{\TPVertModule}{1em}
\begin{textblock}{9}(45,-40)
HUPD-2409
\end{textblock}
\vspace*{-3.4em} % finetunig


\section{Introduction}
\label{introduction}
In high-energy non-central heavy-ion collisions, it is predicted that extremely intense magnetic fields are generated by charged particles moving nearly at the speed of light~\cite{skokov2009estimate,voronyuk2011electromagnetic,PhysRevC.85.044907}.
The intensity of the fields is estimated to reach $|B| \sim 10^{14}$~T (equivalently $|eB| \sim m_{\pi}^{2}$) at the Relativistic Heavy Ion Collider (RHIC) at Brookhaven National Laboratory~\cite{bzdak2012event} and $|eB| \sim 10m_{\pi}^{2}$ at the Large Hadron Collider (LHC) at CERN~\cite{zhong2014systematic, zhong2015spatial} in the early stages of the collisions.
The presence of intense magnetic fields beyond the electron's critical magnetic field $4.4 \times 10^{9}$~T,
can alter the structure of the quantum electrodynamics (QED) vacuum,
potentially leading to non-linear phenomena such as electron pair production via the Schwinger mechanism~\cite{schwinger1951gauge} and photon splitting~\cite{PhotonSplitting}.
Additionally, the magnetic field can also affect the quantum chromodynamics (QCD) vacuum and QCD matter, 
resulting in various physical phenomena,
including a reduction in crossover temperature~\cite{Bali_2012},
the chiral magnetic effect~\cite{CME},
and quark synchrotron radiation~\cite{tuchin2013particleproductionstrongelectromagnetic}.
Thus, these intense magnetic fields are of significant interest in both QED and QCD theoretical discussions.
For recent reviews for the physics related to intense electric/magnetic fields,
see Refs.~\cite{HATTORI2023104068,FEDOTOV20231} and references therein.

Photon vacuum polarization is a fundamental QED effect, where photons interact with the vacuum, altering its structure in the presence of intense electromagnetic fields.
Extensive theoretical studies have been conducted on the vacuum polarization tensor in intense magnetic fields.
We focus on the phenomenon where virtual photons, polarized due to an intense magnetic field, decay anisotropically into lepton pairs relative to the field direction.
This effect, referred to as virtual photon polarization, offers a probe for detecting the magnetic field.

Crucially, the polarization has been shown to depend on the magnetic field intensity~\cite{hattori2013vacuum,ishikawa2013numerical}, making the time evolution of the magnetic field an important factor.
In high-energy heavy-ion collisions, the magnetic field generated by the spectator protons rapidly decays on $\tau \sim \mathcal{O}(1)~\text{fm}/\text{c}$ as the spectators quickly move away~\cite{PhysRevC.85.044907,hattori_novel_2017,bzdak_event-by-event_2012}.
After the formation of the quark-gluon plasma (QGP),
currents within the medium may contribute to the magnetic field, prolonging its lifetime~\cite{gursoy_magnetohydrodynamics_2014}.
Various phenomenological models,
particularly relativistic magnetohydrodynamics (RMHD)~\cite{gursoy_charge-dependent_2018,inghirami_magnetic_2020,inghirami_numerical_2016},
have been employed to calculate the time evolution of the magnetic fields in heavy-ion collisions.
We use relativistic resistive magnetohydrodynamics (RRMHD) which is a more realistic model to compute the time evolution of the magnetic field.
This model accounts for QGP's finite electrical conductivity,
which has been calculated to be non-zero using Lattice QCD, perturbative QCD, and using AdS/CFT~\cite{aarts_electrical_2021,hattori_electrical_2016,li_conductivities_2018}.
Incorporating finite, non-zero conductivity provides a more realistic understanding of the magnetic field's spatiotemporal behavior, which is a crucial parameter for calculating the polarization.

We study the effect of the intense magnetic field through the virtual photon polarization in terms of the asymmetry of muon pair production rate against the expected direction of the magnetic field in the non-central heavy ion collision with the ALCIE detector at CERN-LHC. A short-lived intense magnetic field is predicted from RRMHD model, leading to an expectation of pronounced effects on the asymmetry of muon pair production decayed from virtual photon.
Based on theoretically evaluated polarization, we assessed the experimental significance by scaling Monte Carlo data to the statistics collected in 2010--2011 and in 2015--2018, and then discussed the significance with Monte Carlo data scaled to the higher statistics expected from the 2023--2026 data collection.
The preliminary result has been presented at 79th Annual Meeting of the Physical Society of Japan~\cite{JPS} and 12th International Conference on Hard and Electromagnetic Probes of High-Energy Nuclear Collisions~\cite{HP24}.

This paper is organized as follows:
In Sec.~\ref{b_intensity}, we introduce the model to calculate the time evolution of the magnetic field in Pb-Pb collisions at $\sqrt{s_{\rm NN}} = 2.76$~TeV using a RRMHD model.
In Sec.~\ref{prod_rate}, we explain the muon pair production rate in the presence of the intense magnetic field.
In Sec.~\ref{result}, we show the results of the numerical calculations and discuss the detectability of the intense magnetic field with the ALICE detector.

\section{Estimation of the magnetic fields time evolution}
\label{b_intensity}
An important parameter of the polarization calculation is the magnetic field intensity.
Even though effects of intense magnetic fields have been investigated in heavy-ion collisions, e.g. light-by-light scattering~\cite{2017} and charged dependent directed flow~\cite{DmesonDirectedflow} proving suggestive evidence, exact details on the intensity and lifetime of the magnetic fields remain open questions.
A challenge is modeling the time evolution of the magnetic fields in a conductive medium, QGP, that is rapidly expanding.
To partly overcome that challenge, we use a 3+1D relativistic resistive magnetohydrodynamics (RRMHD) model to estimate the time evolution of the magnetic fields \cite{Nakamura_2023}.
Reviewing the relevant details of that phenomenological model, RRMHD starts with the conservation laws of ideal-hydrodynamics
\begin{align}
    \nabla_\mu N^\mu & = 0,
    \\
    \nabla_\mu T^{\mu\nu} & = 0,
\end{align}
where $N^{\mu}$ denotes the fluid current density, $T^{\mu\nu}$ denotes the energy-momentum tensor,
and $\nabla_\mu$ represents the covariant derivative.
Then those ideal-hydrodynamic equations are augmented with Maxwell's equations, 
\begin{align}
    \nabla_\mu F^{\mu\nu} & = - J^\nu,
    \\
    \frac{1}{2}\nabla_\mu \epsilon^{\mu\nu\rho\sigma} F_{\rho\sigma} & = 0,
\end{align}
where $F^{\mu\nu}$ denotes the magnetic field intensity tensor, $J^{\mu}$ denotes the electric current density, and $\epsilon^{\mu\nu\rho\sigma}$ is the Levi-Civita tensor density.
Finally, Ohm's law is used to close the system of equations,
\begin{equation}\label{eq:ohmslaw}
    J^\mu = q u^\mu + \sigma_{e} F^{\mu\nu}u_\nu.
\end{equation}
We have assumed the electric conductivity $\sigma_{e}$ is a constant and $q=-J^\mu u_\mu$ is the electric charge density in the fluid's comoving frame denoted by the four-velocity $u^{\mu}$.

Although other simpler models can simulate the time evolution of QGP and the electromagnetic fields, they focus on describing only the conductivity, or the rapid expansion of QGP~\cite{inghirami_numerical_2016,inghirami_magnetic_2020,ini_B_tuchin}.
By using a RRMHD model we can calculate the magnetic field intensity and lifetime with the finite electric 
conductivity and expansion of QGP in a self-consistent manner.

The initial energy density and magnetic fields are calculated by using the optical Glauber model and the boosted coulomb fields of the colliding nuclei.
We assume the nuclei only move along the beam-axis, in other words the $z$-axis, and are Lorentz contracted onto the transverse $x$-$y$ plane.
Because we start the hydrodynamic calculation after medium formation, the magnetic fields are produced in a conductive medium.
We calculate the initial magnetic fields including a scalar electric conductivity $\sigma_{e}$ to describe the medium as in Ref.~\cite{Nakamura_2023} that is based on Ref.~\cite{ini_B_tuchin}.

This method for calculating the initial magnetic fields simplifies the true pre-equilibrium dynamics of the collision (see \cite{gursoy_charge-dependent_2018} for a detailed discussion on the difficulities of the pre-equilibrium EM fields).
Thus, we neglect the contribution of polarization from the magnetic field in the pre-equilibrium phase.

% \section{Calculation method of virtual photon polarization}
\section{Muon pair production asymmetry in the intense magnetic field}
\label{prod_rate}
Experimentally virutal photon polarization due to an intense magnetic field can emerge as an anisotropy of the production rate of a muon pair from a virtual photon;
because  the propagation of photons follows an anisotropic response due to the virtual photon polarization relative to the direction of the magnetic field at the one-loop level~\cite{ishikawa2013numerical}.
We focus on this source as the anisotropy of the muon pair production ratio from virtual photons.
The production rate is given by:
\begin{equation}
   R_{\mu^{+}\mu^{-}}=\frac{\alpha^2}{2\pi^4}\mathrm{L}^{\mu\nu}(p_1,p_2)
\mathrm{D}_{\mu\alpha}(q,eB)\mathrm{D}^*_{\nu\beta}(q,eB)\frac{\mathrm{ImG}^{\alpha\beta}_R(q,T,eB)}{\mathrm{e}^{q^0/T}-1},
   \label{eq:DimuonProdRate}
\end{equation}
where $\alpha$ is the fine-structure constant, $e$ is the electric charge, $B$ denotes the magnitude of the magnetic field; $B=\abs{\rm{B}}$, and $T$ represents the temperature of the system.
$p_1$ and $p_2$ are the muon and anti-muon four-momentum, $q$ is the virtual photon momentum
satisfying $q = p_1 + p_2$.
$\mathrm{L}^{\mu\nu}$ is the leptonic tensor given by
\begin{equation}
   \mathrm{L}^{\mu\nu}=p^{\mu}_1p^{\nu}_2+p^{\nu}_1p^{\mu}_2-\qty(p_{1}\cdot p_{2}+m_\mu^2)g^{\mu\nu}.
   \label{eq:LepTensor}
\end{equation}
The virtual photon source term $\textrm{ImG}^{\alpha\beta}_R(q,T,eB)/(\mathrm{e}^{q^0/T}-1)$ 
contains the retarded Green function of the electric currents in the thermal and magnetic fields.
Since it is difficult to show the details of the virtual photon source
in high-energy heavy-ion collision, we replace it with the current conservation form as
\begin{equation}
   \frac{\mathrm{ImG}^{\alpha\beta}_R(q,T,eB)}{\mathrm{e}^{q^0/T}-1}=\qty(-g^{\alpha\beta}q^2+q^{\alpha}q^{\beta})C,
   \label{eq:source}
\end{equation}
where $C$ is a constant.
The effect of the polarization of the photon source term is partially canceled in the polarization ratio
for the anisotropy. The photon propagator $\textrm{D}_{\mu\nu}(q,eB)$ is given by
\begin{equation}
   \mathrm{D}_{\mu\nu}(q,eB)=-\frac{i}{q^2}\qty[g^{\mu\nu}-\frac{1}{q^2}\Pi^{\mu\nu}(q,eB)]^{-1},
   \label{eq:propagator}
\end{equation}
where $\qty[A^{\mu\nu}]^{-1}$ means the matrix inverse of the tensor $A^{\mu\nu}$.
The vacuum polarization tensor $\Pi^{\mu\nu}$ is evaluated at one-loop level in the Furry picture
with a constant intense magnetic field, and evaluated numerically in the Landau summation form ~\cite{Hattori:2012je,ishikawa2013numerical}.
We consider the contribution of electron and muon for virtual fermions included in the polarization tensor.

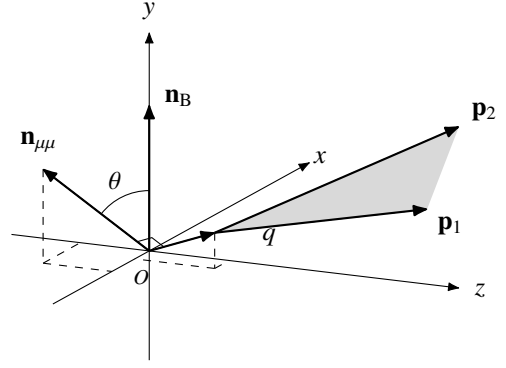
\begin{figure}
    \centering
    \tdplotsetmaincoords{75}{115}
    \begin{tikzpicture}[tdplot_main_coords]
        \draw[->, >={Latex[round]}] (3,0,0) -- (-5,0,0);
        \node at (-5.3,0,0) {$x$};
        \draw[->, >={Latex[round]}] (0,-2,0) -- (0,4.5,0);
        \node at (0,4.8,0) {$z$};
        \draw[->, >={Latex[round]}] (0,0,-1.5) -- (0,0,3);
        \node at (0,0,3.3) {$y$};
        \node at (0.25,0,-0.3) {\fontsize{8pt}{10pt}\selectfont $O$};

        \coordinate (O) at (0, 0, 0);
        \coordinate (p1) at (2.50, 5.20, 1.77);
        \coordinate (p2) at (1.51, 5.20, 2.66);
        \coordinate (q) at (0.45, 1.16, 0.49);
        \coordinate (qxz) at (0.45, 1.16, 0);
        \coordinate (qx) at (0.45, 0, 0);
        \coordinate (qz) at (0, 1.16, 0);
        \coordinate (B) at (0, 0, 2);
        \coordinate (dimuon) at (1.16, -1.00, 1.29);

        \draw[thick,->, >={Latex[round]}] (0,0,0) -- (B);
        \node at (-0.5, 0.2 , 2) {$\mathbf{n}_{\rm{B}}$};
        
        \draw[thick, ->, >={Latex[round]}] (0, 0, 0) -- (q);
        \node at (-0.3, 1.6, 0.3) {$q$};
        \draw [dashed] (p1) -- (q);
        \draw [dashed] (p2) -- (q);
        
        \fill[gray!30] (q) -- (p1) -- (p2) --cycle;
        \draw[thick,->, >={Latex[round]}] (q) -- (p1);
        \node at (2.2, 5.4, 1.5) {$\mathbf{p}_{1}$};
        \draw[thick,->, >={Latex[round]}] (q) -- (p2);
        \node at (1.14, 5.4, 2.80) {$\mathbf{p}_{2}$};
    
        \draw[thick,->, >={Latex[round]}] (0,0,0) -- (dimuon);
        \draw[dashed] (dimuon) -- (1.16, -1.00,  0.0);
        \draw[dashed] (1.16, -1.00,  0.0) -- (1.16 ,  0.0 ,  0.0);
        \draw[dashed] (1.16, -1.00,  0.0) -- (0.0 ,  -1.0 ,  0.0);
        \node at (1.3,-1,1.7){$\mathbf{n}_{\mu\mu}$};
    
        \draw (dimuon) -- (0,0,0) coordinate (O) -- (B);
        \draw pic[draw=black, "$\theta$",angle eccentricity=1.3, angle radius=0.8cm] {angle=B--O--dimuon};

        \draw (q) -- (0,0,0) coordinate (O) -- (dimuon);
        \draw pic[draw=black, angle radius=0.2cm]
        {right angle=q--O--dimuon};

        \draw [dashed] (q) -- (qxz);
        \draw [dashed] (qxz) -- (qz);
        \draw [dashed] (qxz) -- (qx);
        
    \end{tikzpicture}
    
    \caption{The definition of the angle between decay plane of muon pair and the intense magnetic field. $\mathbf{p}_{1}$ and $\mathbf{p}_{2}$ are momentum of $\mu^{-}$ and $\mu^{+}$.The vector $\mathbf{n}_{\rm{B}}$ is the unit vector of $\mathbf{B}$, and $\mathbf{n}_{\mu\mu}$ is the unit vector of the decay plane of muon pair which is determined by $\mathbf{p}_{1}$ and $\mathbf{p}_{2}$.}
    \label{fig:decayplane}
\end{figure}

To define the virtual photon polarization, we employ the coordinate system depicted in Fig.~\ref{fig:decayplane}.
The momenta, $\mathbf{p}_{1}$ and $\mathbf{p}_{2}$, are momentum of muon and anti-muon, respectively.
The normal vector of the decay plane of the muon pair $\mathbf{n}_{\mu\mu}$ is defined by
$\mathbf{n}_{\mu\mu} = \mathbf{p}_{1}\times\mathbf{p}_{2}/|\mathbf{p}_{1}\times\mathbf{p}_{2}|$.
The angle $\theta$ between the constant magnetic field $\mathbf{B}$ and the normal vector of the decay plane is then determined by $\theta=\cos^{-1}(\mathbf{n}_{\mu\mu}\cdot\mathbf{n}_{\rm{B}})$, where 
$\mathbf{n}_{\rm{B}}$ is the unit vector of $\mathbf{B}$ that defines the $y$-direction of the coordinate system.
It is expected that the direction of the magnetic field is perpendicular to the event plane of the non-central collisions in high-energy heavy-ion collision experiments,
by which we can relate the $y$-direction of the coordinate system to the experimental data.
We set that the $z$-direction corresponds to the beam axis of the heavy ion collision experiment, though the production rate has the rotation symmetry in the $x$--$z$ plane.

Using that coordinate system, we define the virtual photon polarization by the muon pair production rate:
\begin{equation}
P_{\mathrm{cal}}=\frac{R_{\perp}-R_{\parallel}}{R_{\perp}+R_{\parallel}},
   \label{eq:calculatedPol}
\end{equation}
where $R_{\perp}$ is the muon pair production rate Eq.~\eqref{eq:DimuonProdRate} evaluated at the pair production perpendicular to the magnetic field which is $\theta=0$ and $R_{\parallel}$ is the rate evaluated at the production parallel to the magnetic field which is $\theta=\pi/2$.

We numerically evaluate the virtual photon polarization Eq.~\eqref{eq:calculatedPol} followed by substituting the magnetic field intensity using the model described in Sec.~\ref{b_intensity}.
The numerical results will be explained in the next section together with the details of the experimental setups, the time evolution of the magnetic field, and the muon kinematics.

\section{Numerical calculation results}
\label{result}
In this section, we show the estimation of the virtual photon polarization and the feasibility of measuring it for Pb-Pb collisions at $\sqrt{s_{\rm NN}} = 2.76~$TeV conducted by the ALICE collaboration based on these results.
We considered the measurement of muon pairs originating from the decay of prompt virtual photons produced in the early stages of the collision.

To evaluate Eq.~\eqref{eq:calculatedPol} numerically we have to determine the geometry and the value of the virtual photon four momentum.
Because during our detailed calculation we observed that $P_{\mathrm{cal}}$ increases with smaller virtual photon mass and the muon pair production requires $q^2 > (210$~MeV/$c^{2})^2$, also a lighter virtual photon is preferred in the experimental yield we employ $q^2=(300$~MeV/$c^{2})^2$ for the virtual photon mass in the following analysis.
Concerning the geometry, the virtual photon can be assumed to propagate along the $z$-axis because the total three-momentum of the muon pair aligns with the beam axis within the acceptance of ALICE's muon detector ($2.5 < |\eta| < 4.0$)~\cite{ALICE:2008ngc}.

To compute $R_{\perp}$ and $R_{\parallel}$ we substitute a magnetic field intensity into Eq.~\eqref{eq:DimuonProdRate}.
Then, these values are used in Eq.~\eqref{eq:calculatedPol} to determine $P_{\mathrm{cal}}$.
In the calculation of the vacuum polarization tensor, we set the upper limits of the Landau level sums to $(n_{\mathrm{max}}, l_{\mathrm{max}}) = (1 \times 10^3, 1 \times 10^4)$, as discussed in Ref.~\cite{ishikawa2013numerical}, to ensure sufficient convergence.

From here, we illustrate the time evolution of the magnetic field.
To evaluate the contribution of the decaying magnetic field during the propagation of virtual photons, it is necessary to take the time evolution of the magnetic field into account.
This is because the magnetic field rapidly decays from its initial value.
Additionally, the initial intensity of the magnetic field depends on the collision centrality because the field are created by the protons that are spectators of the collision.
To maximize the magnetic field intensity we calculate it for a mid-central collision using the impact parameter $b=10$~fm, at which hydrodynamics can still be applied.

\begin{figure}
    \includegraphics[width=1.0\linewidth]{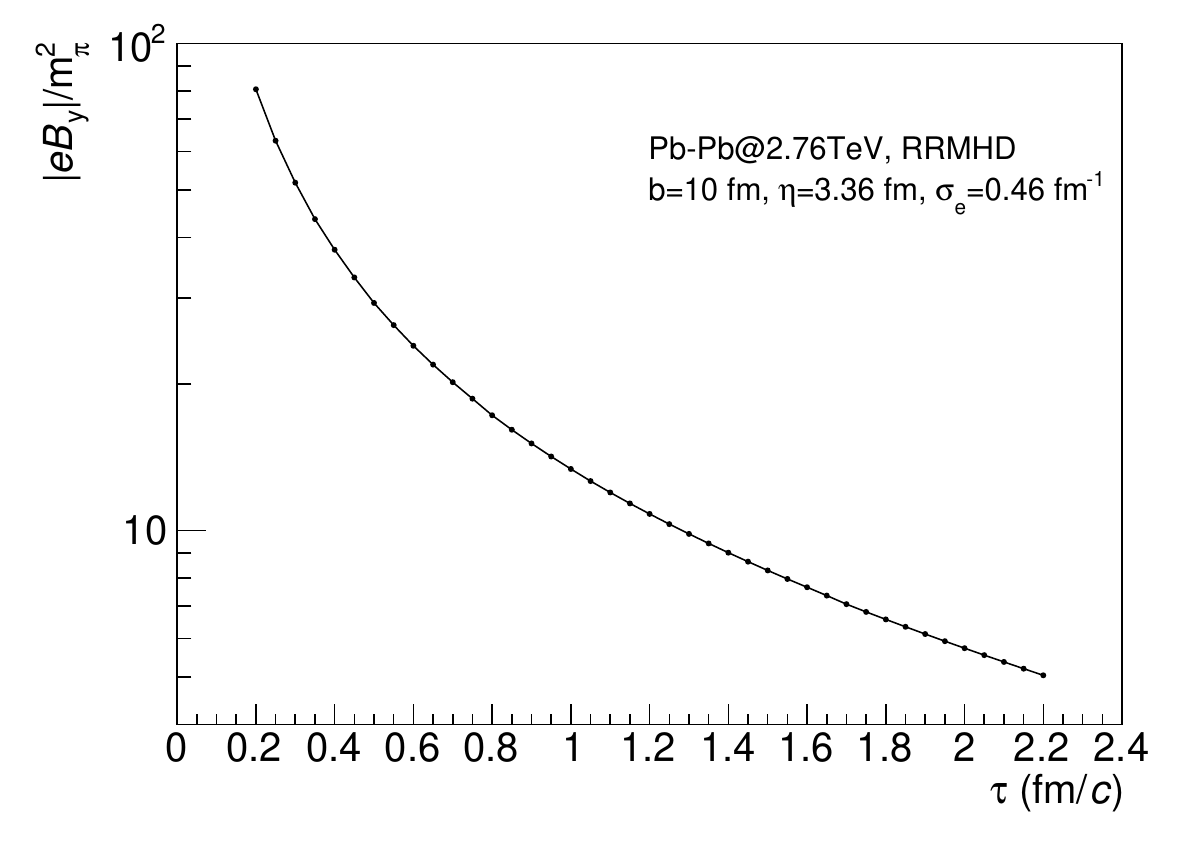}
    \centering
    \caption{The time evolution of the magnetic field at the center of the collision using the model described in \cite{Nakamura_2023}.
    The impact parameter $b=10$ fm was set to maximize the initial field value and the conductivity $\sigma_{e} = 0.46$ \textrm{fm}$^{-1}$ was chosen to be larger than the lattice result often quoted in the literature.
    This larger conductivity was chosen because separate calculations show the model under predicts experimental charged directed flow values when the lattice conductivity is used.}
    \label{fig:Bintensity}
\end{figure}

The RRMHD model is initialized at $\tau = 0.2$~fm/$\it{c}$ with the parameters of table 1 in Ref.~\cite{inghirami_magnetic_2020}\footnote{We separately checked the consistency of those parameters for our model and found it can describe the experimental results well.}.
We chose an electric conductivity for QGP independently of lattice and pQCD results.
Instead, we use the electric conductivity $\sigma_{e} = 0.46~\text{fm}^{-1}$ that is about an order of magnitude larger than what is often quoted in literature.\footnote{Although this larger value does increase the later time magnetic field intensity, we have found our RRMHD model generally under-predicts the intensity of the EM fields.
This is because the spectator nuclei contribution to the EM currents are left out during the time evolution.
The details on why that occurs and how this value was chosen will be reported separately as that is not the focus of this work.}
We extracted the center of grid value for $B_y$ inside the freezeout region where the fluids energy density exceeds 0.15~$\text{GeV}/\text{fm}^{3}$ with a time step 0.1~fm/$c$.
The result of the RRMHD model for the time evolution of the magnetic field is plotted in Fig.~\ref{fig:Bintensity}.

\begin{figure}
    \centering
    \includegraphics[width=1.0\linewidth]{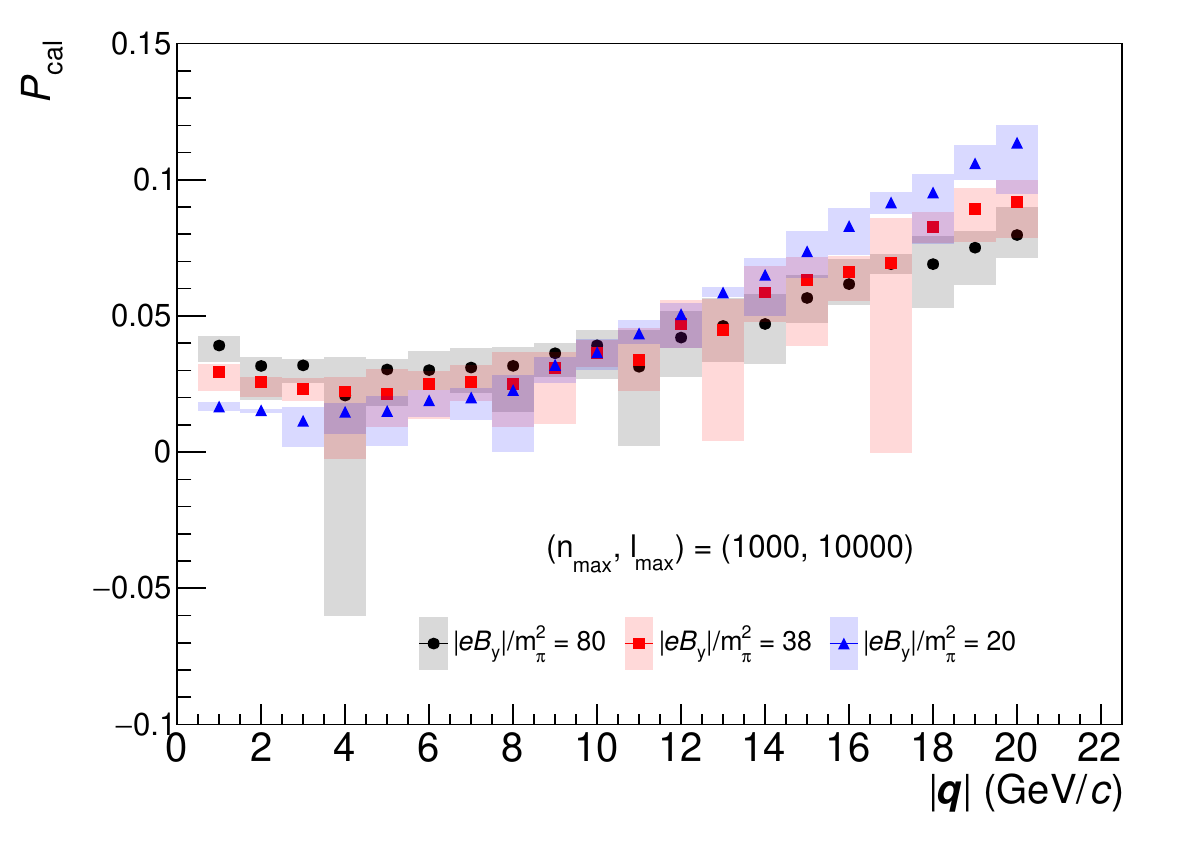}
    \caption{The virtual photon momentum dependence of the virtual photon polarization at $|eB_{\mathrm{y}}|/m^{2}_{\pi}$ = 80, 35 and 20.}
    \label{fig:momentum_dep}
\end{figure}

Next, we display the calculation results of $P_{\mathrm{cal}}$ with the calculated magnetic field intensity at each time step.
Figure~\ref{fig:momentum_dep} shows $P_{\mathrm{cal}}$ as a function of the magnitude of 3-momentum of the virtual photon $|\mathbf{q}|$ for magnetic field intensity of $|eB_{\mathrm{y}}|/m^{2}_{\pi} =$ 80, 35 and 20.
The systematic uncertainty shown in Fig.~\ref{fig:momentum_dep} originates from the experimental resolution of the virtual photon momentum and fluctuations due to the threshold where the imaginary component of the vacuum polarization tensor emerges as discussed in Ref.~\cite{ishikawa2013numerical}.
This threshold is defined by $s^{ln}_{+} \equiv 1/4(\sqrt{1+2l\mu} + \sqrt{1+2(l+n)\mu})^{2} < r$, and it depends on the Landau level numbers $l~\text{and}~n$, photon energy $r=k_{\parallel}/4m^{2}$, and magnetic field intensity $\mu=|eB_{\mathrm{y}}|/m^{2}$ and at the threshold $P_{\mathrm{cal}}$ exhibits discontinuities and spikes~\cite{ishikawa2013numerical}.
We evaluated this uncertainty by smearing the momentum at each bin. The uncertainty of momentum in the measurement was assumed to follow a Gaussian distribution with a standard deviation of $0.25 \, \mathrm{GeV}/c$.
Based on this Gaussian distribution, 10 random samples were generated for each momentum, and the polarization was calculated for each sample.
The averaged value of the 10 samples is plotted with the uncertainty determined by the maximum and minimum values among the 10 samples at each momentum bin.
As shown in the figure, $P_{\mathrm{cal}}$ increases with momentum for all magnetic field intensity, indicating that measurements at higher momentum are preferable.

\begin{figure*}[t]
    \centering
    \includegraphics[width=1.0\linewidth]{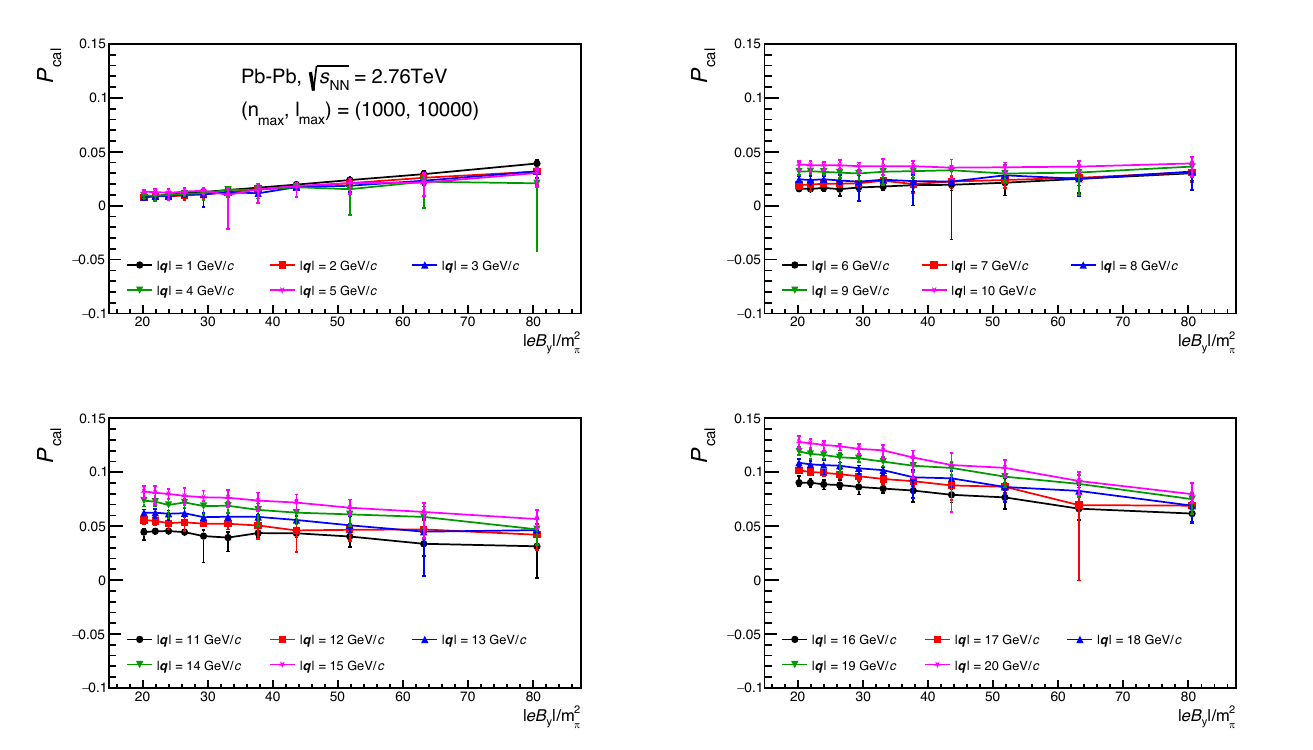} 
    \caption{The magnetic field intensity dependence of the virtual photon polarization in the momentum range $1<|\mathbf{q}|<20$ GeV/$c$ and the time range $20 < \abs{eB_y}/m_{\pi}^2 < 80$ in Pb-Pb collisions at $\sqrt{s_{\mathrm{NN}}}=2.76$ TeV. The calculations were performed using the magnetic field intensity at each time step from Fig.~\ref{fig:Bintensity}. This means that a stronger magnetic field corresponds to an earlier time for the virtual photon polarization.}
    \label{fig:Bintensity_dep}
\end{figure*}

We now discuss the expected $P_{\mathrm{cal}}$ with the ALICE detector.
Figure~\ref{fig:Bintensity_dep} shows the magnetic field intensity dependence of $P_{\mathrm{cal}}$ for several virtual photon momenta in the range $1<|\mathbf{q}|<20$~GeV/$c$, calculated by using the magnetic field intensity at each time step as computed in Fig.~\ref{fig:Bintensity}. 

To evaluate the expected $P_{\mathrm{cal}}$ with the ALICE detector, we have to perform an averaging processes on the theoretical $\expval{P_{\mathrm{cal}}}$ according to detector limitations and a realistic initial virtual photon production rate.
Experimentally the time resolution is not precise enough to detect rapid changes in the muon pair production rate, so we have to average over a time range for the prompt virtual photon decays.
To fix the time range for prompt photon production, we employ $0.2<\tau<0.7~$fm/$c$ from the lifetime of prompt virtual photons estimated to be $\tau>\hbar/(M_{\gamma^{*}}c^{2})\simeq0.7~$fm/$c$ based on the uncertainty principle.
Thus we average over the polarization between the magnetic field values $20 <\abs{eB_y}/m_{\pi}^2<80$ at each $|\mathbf{q}|$ based on Figs. \ref{fig:Bintensity} and \ref{fig:Bintensity_dep}.
When $P_{\mathrm{cal}}$ is averaged over $\abs{\mathbf{q}}$ it requires the virtual photon production rate because the constant in Eq.~\eqref{eq:source} would depend on $\abs{\mathbf{q}}$.
So, we employ a $\abs{\mathbf{q}}$ dependent virtual photon production rate obtained by combining the Kroll-Wada formula, which describes the relationship between the yield of real photons and the production of muon pairs via virtual photon interactions~\cite{KrollWada}, and the pQCD formula for the yield of prompt real photons~\cite{PhysRevC.93.044906}.
The integration range over $\abs{\mathbf{q}}$ is $1 <\abs{\mathbf{q}}< 20$~GeV/$c$, which is estimated based on the experimental conditions that the total transverse momentum of a muon pair $p_T$, equivalently the transverse momentum of a virtual photon, is detected in the range $0.1 < p_T < 2$~GeV/$c$ and the geometrical orientation of the rapidity region for the ALICE detector.
As a result we obtain
\begin{align}
\expval{P_{\mathrm{cal}}} = 0.05,  
\label{eq:avePcal}
\end{align}
after averaging over $20 < \abs{eB_y}/m_{\pi}^2 < 80$ and $1 < \abs{\mathbf{q}} < 20$ GeV/$c$ as described above.
Additionally, since $P_{\mathrm{cal}}$ increases with $\abs{\mathbf{q}}$ as shown in Fig.~\ref{fig:momentum_dep}, the value of $P_{\mathrm{cal}}$ for a wider range should increase.
Thus we consider $\expval{P_{\mathrm{cal}}}=0.05$ as the lower bound for the averaged polarization.
\footnote{
A larger $\expval{P_{\mathrm{cal}}}$ in the higher $p_T$ range is preferable to increase the detectability of the polarization experimentally.
However, due to our limitation on the Landau level summation formula of the virtual photon propagator in the high $q$ region, it is difficult give a concrete value of the averaged polarization in higher  $\abs{\mathbf{q}}$.}

We will now evaluate the significance of the photon polarization detectability at $\sqrt{s_{\rm NN}} = 2.76~$~TeV with the ALICE detector.
Since prompt photons become dominant over direct photons generated through hadron decays in $p_{\text{T}} > 4$~GeV/$c$~\cite{PhysRevC.93.044906}, it is experimentally reasonable to measure the transverse momentum of muon pairs within this range.
To evaluate the significance, although Eq.~\eqref{eq:avePcal} was evaluated for muon pairs with $0.1<p_{\text{T}}<2~$GeV/$c$, we employ $\expval{P_{\mathrm{cal}}}=0.05$ for muon pairs with $p_{\text{T}} > 4$~GeV/$c$.
The statistical significance $\sigma$ is defined by 
\begin{align}
\sigma=\dfrac{P_{\mathrm{meas}}}{\Delta P_{\mathrm{meas}}},
\end{align}
where $P_{\mathrm{meas}}$ is experimentally measured virtual photon polarization and 
$\Delta P_{\mathrm{meas}}$ represents the statistical uncertainty.
We estimated $P_{\mathrm{meas}}$ by accounting for the yields of both muon pairs from prompt virtual photon as signal and the background signal:
\begin{align}
P_{\mathrm{meas}}&=\dfrac{N_{\perp}-N_{\parallel}}{N_{\perp}+N_{\parallel}},\\
N_{\perp}&=\dfrac{N_{\mathrm{BG}}}{2}+\dfrac{1+\expval{P_{\mathrm{cal}}}}{2}N_{\mathrm{S}},\\
N_{\parallel}&=\dfrac{N_{\mathrm{BG}}}{2}+\dfrac{1-\expval{P_{\mathrm{cal}}}}{2}N_{\mathrm{S}},\\
\Delta P_{\mathrm{meas}} &= \dfrac{2}{N_{\perp}+N_{\parallel}}\sqrt{\dfrac{N_{\perp}N_{\parallel}}{N_{\perp}+N_{\parallel}}},
\end{align}
where $\expval{P_{\mathrm{cal}}}=0.05$ and $N_{\mathrm{S}}$ and $N_{\mathrm{BG}}$ represent
the signal ($\gamma^*\to\mu\mu$) event counts and the background event counts, respectively.
We evaluated $N_{\mathrm{S}}$ by simulating pp collisions with PYTHIA 8~\cite{bierlich2022comprehensiveguidephysicsusage, Skands_2014} followed by scaling the results to the statistics in Pb-Pb collisions collected between 2010 and 2011 with the ALICE detector~\cite{2014}.
While $N_{\mathrm{BG}}$ was calculated directly by simulating Pb-Pb collisions with PYTHIA 8~\cite{Bierlich_2018}.
For the yield of the background signal $N_{\mathrm{BG}}$, we counted combinatorial muon pairs arising from random pairings of muons in the data set.
We found that the statistical significance was $0.07\sigma$ for $P_{\mathrm{meas}}$ deviating from zero.
The data collected with the ALICE detector at $\sqrt{s_{\mathrm{NN}}} = 5.02~$TeV in 2015--2018 increased the statistics by about 5 times compared to the 2010--2011 data~\cite{journeyQCD}, leading to a modest improvement in significance to $0.15\sigma$ assuming $\expval{P_{\mathrm{cal}}} = 0.05$.
These are a rather low significance because of low statistics accumulated at these runs.
Fortunately the statistics will be extremely improved in the on-going and future experiments at other center of mass energies conducted by ALICE\@.

The ongoing Pb-Pb collision experiment at $\sqrt{s_{\rm NN}} = 5.36~$TeV, planned to run between 2023 and 2026~\cite{2014,Acharya_2024}, will collect a significantly larger amount of data, which is expected to be about hundred times of those collected in 2010--2011 and 2015--2018, and the new data will improve the significance of the signal of the photon polarization.
The experimental improvements for the high statistics are accomplished by an improved collision rate made by the upgraded accelerator, 
and a new data processing system enabling continuous data readout with the upgraded ALICE detector; details of this upgrade can be found in \cite{Acharya_2024}. 
We also evaluated the significance of the detection of the virtual photon polarization with the expected amount 
of data at $\sqrt{s_{\rm NN}} = 5.36$~TeV as have been done for data at earlier runs.
We obtained a significance of $\sim 1.7\sigma$ assuming $\expval{P_{\mathrm{cal}}} = 0.05$.\footnote{
Although, at $\sqrt{s_{\rm NN}} = 5.36$~TeV, $\expval{P_{\mathrm{cal}}}$ could be slightly improved due to a more pronounced magnetic field intensity because of a larger Lorentz factor $\gamma$~\cite{ini_B_tuchin}, 
we use $\expval{P_{\mathrm{cal}}} = 0.05$ for a conservative estimate.
}
This demonstrates that through the ongoing and future ALICE runs, virtual photon polarization could become a promising probe of the intense magnetic fields produced in high-energy heavy-ion collisions.

\section{Summary}
%%\label{}
We numerically calculated the virtual photon polarization for the first time and evaluated the detectability of the intense magnetic fields that are generated in high-energy heavy-ion collisions using the polarization of muon pairs from the decay of virtual photons, which we proposed as a new probe with the ALICE detector at CERN-LHC.
The muon pair from the decay of prompt virtual photons is a clean probe, unaffected by the QGP, which is dominated by strong interactions, allowing direct access to the initial magnetic field.
The polarization depends on the magnetic field intensity, which decreases during the propagation of the virtual photons, making the time evolution of the magnetic field a key factor.
We utilized a RRMHD model that incorporates a finite electrical conductivity to compute the time evolution of the magnetic field.
Then, we calculated the vacuum polarization tensor in the presence of strong magnetic fields and evaluated the virtual photon polarization within $1 < \abs{\mathbf{q}} < 20$~GeV/$c$ and the time $0.2 < \tau < 0.7$ fm/$c$.
We estimated the averaged virtual photon polarization and obtained $\expval{P_{\mathrm{cal}}}=0.05$.
The experimental significance is evaluated by incorporating the yield of prompt virtual photon and background in Pb-Pb collisions, using PYTHIA8, for the existing data at $\sqrt{s_{\mathbf{NN}}}=2.76$~TeV and 5.02~TeV
and for the ongoing run at $\sqrt{s_{\mathbf{NN}}}=5.36$~TeV between 2023 and 2026.
We found the statistical significance of the polarization is estimated to be $0.07\sigma$ and $0.15\sigma$
with the data statistics of Pb-Pb collisions at $\sqrt{s_{\mathbf{NN}}}=2.76$~TeV in 2010--2011 and at 5.02~TeV in 2015--2018, respectively,
but should improve to $\sim 1.7\sigma$ for the new data at $\sqrt{s_{\mathbf{NN}}}=5.36$~TeV\@.
This indicates that the virtual photon polarization method can be an effective probe for detecting the intense magnetic fields with the ALICE detector\@.

\section*{Acknowledgements}
We thank Koichi Hattori for useful discussions.
The numerical calculations have been done with the PC cluster at Hiroshima University (KK). 
This work is supported in part by Hiroshima University Graduate School Research Fellowship (KK) and by the World Premier International Research Center
Initiative (WPI) under MEXT, Japan (CN and KS)
and by Japan Society for the Promotion of Science
(JSPS) KAKENHI Grant Nos. JP20H00156, JP20H11581 (CN), JP18H05401, JP20H00163 (KS). 
Numerical computation of the RRMHD model in this work was carried out at the Yukawa Institute Computer Facility.

% \appendix

% \section{Appendix title 1}
% %% \label{}

% \section{Appendix title 2}
% %% \label{}

\bibliographystyle{elsarticle-num}
\bibliography{example}



%% For example:

%% \bibitem[Aladro et al.(2015)]{Aladro15} Aladro, R., Martín, S., Riquelme, D., et al. 2015, \aas, 579, A101

%%\end{thebibliography}

\end{document}